# Experimental Signatures of Spin Superfluid Ground State in Canted Antiferromagnet $Cr_2O_3$ via Nonlocal Spin Transport

Wei Yuan[1,2], Qiong Zhu[1,2], Tang Su[1,2], Yunyan Yao[1,2], Wenyu Xing[1,2], Yangyang Chen[1,2], Yang Ma[1,2], Xi Lin[1,2], Jing Shi[3]*, Ryuichi Shindou[1,2], X. C. Xie[1,2]*, and Wei Han[1,2]*

[1]International Center for Quantum Materials, School of Physics, Peking University, Beijing 100871, P. R. China

[2]Collaborative Innovation Center of Quantum Matter, Beijing 100871, P. R. China

[3]Department of Physics and Astronomy, University of California, Riverside, California 92521, USA

*Correspondence to: weihan@pku.edu.cn (W.H.); jing.shi@ucr.edu (J.S.); and xcxie@pku.edu.cn (X.C.X.).

**Abstract**: Spin superfluid is a novel emerging quantum matter arising from the Bose–Einstein condensate (BEC) of spin-1 bosons. We demonstrate the spin superfluid ground state in canted antiferromagnetic $Cr_2O_3$ thin film at low temperatures via nonlocal spin transport. A large enhancement of the nonlocal spin signal is observed below ~ 20 K, and it saturates from ~ 5 K down to 2 K. We show that the spins can propagate over very long distances (~ 20 μm) in such spin superfluid ground state and the nonlocal spin signal decreases very slowly as the spacing increases with an inverse relationship, which is consistent with theoretical prediction. Furthermore, spin superfluidity has been investigated in the canted antiferromagnetic phase of the $(11\bar{2}0)$-oriented $Cr_2O_3$ film, where the magnetic field dependence of the associated critical temperature follows a 2/3 power law near the critical point. The experimental demonstration of the spin superfluid ground state in canted antiferromagnet will be extremely important for the fundamental physics on the BEC of spin-1 bosons and paves the way for future spin supercurrent devices, such as spin-Josephson junctions.

**One Sentence Summary:** We report the direct experimental signatures for the spin superfluid ground state in canted antiferromagnet via nonlocal spin transport.

**Main Text:**

The Bose–Einstein condensate (BEC) refers to the quantum state of matter when a large fraction of bosons occupy in the lowest accessible quantum state and has been observed in liquid helium ($^4$He), cold gases, polaritons, and photons, etc (*1-6*). Recently, this concept has been much more general; for example, the BEC of spin-1 bosons and the spin superfluid (*7-14*). To achieve this spin superfluid state, a great deal of effort has been made. For example, microwave pumping has been used to generate the BEC of magnons and supercurrent at room temperature (*9, 14*). However, since the condensed magnons are inherently thermal excitations, the spin superfluidity is not the ground state, and thus occurs only in a very short time scale (micro seconds). Theoretically, the spin transport in spin superfluid has been recently proposed to for ferromagnetic graphene and the ν = 0 quantum Hall state of graphene (*15-17*), ferromagnetic materials (*18-20*) and antiferromagnetic insulator (*21-23*). Besides, recent progresses of BEC in quantum magnets exhibit a thermodynamic ground state at low temperatures inferred from magnetization and heat capacity measurements (*7, 8, 10, 11, 13*), where the condensing spin-1 bosons can be regarded as the mapping of the spin ground state onto a lattice of bosons (*13*). However, despite the enormous effort in this field, the direct experimental observation of real spin transport in the spin superfluid ground state has been lacking.

We report the experimental observation of long distance spin transport in the spin superfluid ground state in canted antiferromagnetic $Cr_2O_3$ thin films at low temperatures, which provides the direct experimental evidence of the spin superfluid ground state from the BEC in canted antiferromagnet. The high quality $Cr_2O_3$ thin films are grown on (0001) - oriented $Al_2O_3$ substrates via pulsed laser deposition (*24*). The spin configurations on the Cr atoms ($Cr^{3+}$, s = 3/2) are illustrated in Fig. 1A, which shows that the spins are aligned along the crystal's [0001] orientation. The spin transport in the spin superfluid state is performed using the nonlocal geometry, as shown in Figs. 1B and 1C inset. The spins are injected into the antiferromagnetic $Cr_2O_3$ thin films from the local Joule heating on the Pt strip via spin Seebeck effect (*25-27*). Then the spin information transports in the spin superfluid state, which can propagate to the right Pt electrode and be detected by voltage measurement across the Pt strip via the inverse spin Hall effect using standard low frequency lock-in technique (*24*). Since the spins are injected by thermal heating via spin Seebeck effect and detected via inverse spin Hall effect, the nonlocal spin signal is probed by measuring the second harmonic nonlocal resistance ($R_{2\varpi}$).

Figure 1C shows the $R_{2\omega}$ as a function of the magnetic field angle ($\varphi$) measured on the nonlocal device on the ~ 19 nm $Cr_2O_3$ thin film with a spacing ($d$) of 10 μm between the two Pt strips. During the measurement, the temperature is 2 K and the in-plane static magnetic field is 9 T. The in-plane magnetic field is used to generate a canted antiferromagnetic phase, as shown in Fig. 1B. The effective spins are detected via the inverse spin Hall effect of Pt and the second harmonic nonlocal resistance is expected to be proportional to sin ($\varphi$) (26):

$$R_{2\omega} = \frac{1}{2} R_{NL} \sin(\varphi) \qquad (1)$$

where the $R_{NL}$ is the nonlocal spin signal. The red solid line is a fitting curve based on equation (1), from which $R_{NL}$ is determined to be 0.79 ± 0.02 V/A². Fig. 2A shows typical representative $R_{2\omega}$ curves on this device ($d$ = 10 μm) as a function of the magnetic field angle ($\varphi$) in 9 T at 2, 5, 10, 15 and 30 K respectively. The nonlocal spin signal as a function of the temperature from 2 to 110 K is summarized in Fig. 2B. No clear nonlocal spin signal is observable when the temperature is higher than 110 K. At ~ 60 K, a modest enhancement of the nonlocal spin signal is observed (24). When the temperature further decreases to ~ 20 K, the nonlocal spin signal starts to increase dramatically. From 20 K to 5 K, a large enhancement of the nonlocal spin signal is observed. Furthermore, the nonlocal spin signal exhibits a saturation feature when the temperature decreases from ~ 5 K down to 2 K. These low temperature behaviors cannot be attributed to the thermally-excited incoherent magnons, the population of which decreases as the temperature decreases below the Neel temperature (29-32).

We attribute the observation of the low temperature enhancement and the saturation of the nonlocal spin signal to the spin superfluid ground state in a canted antiferromagnet, as depicted in Fig. 1B. In such spin superfluid ground state, the spin-1 bosons exhibit the Bose-Einstein condensation at low temperatures, as discussed previously in the content of BEC in quantum magnets (13). This spin superfluid ground state is totally different from the magnon BEC, where the magnons are inherently excitations and have a finite lifetime, and thereby the spin supercurrent and magnon BEC only occur in very short time scales (9, 14). While the BEC in our case is a thermodynamic ground state, for which the spin Hamiltonian originally has the U(1) spin-rotational symmetry around the magnetic field direction (x-direction). On lowering temperature, the U(1) symmetry is spontaneously broken by the canted antiferromagnetic order, where a free U(1) processional motion of the antiferromagnetic moment within the *y-z* plane leads to a spin supercurrent of the *x*-component of the spin density (Fig. 1B). In other words, the spin component in the *y-z* plane ($S_y + iS_z$) acquires a phase rigidity in the spin superfluid state, which results in a spin analog of the superconducting effect. Despite that the injected magnons by the thermal method are usually incoherent, a BEC condensation from incoherent magnons could happen via magnon scatterings (*9, 33, 34*). The detailed condensation of the incoherent magnons into the spin superfluid state needs future studies. The nonlocal spin signal vs. the inverse of temperature is plotted in Fig. 2B inset. Furthermore, this saturation feature below ~ 5 K agrees well with the theoretical calculation of nonlocal spin signal in the spin superfluid ground state below a critical temperature (*15*).

Theoretically, the spin superfluid density is expected to show no variation, or very slow decaying in the presence of the spin-orbit interaction or magnetic damping (*13, 15, 22*). To verify this property, we study the nonlocal spin transport as a function of the spacing between the two Pt strips. To avoid the thermal effect close to the spin injector (*35*), we purposely vary the spacing from 2 μm to 20 μm that is significantly longer than the width of the Pt strip (300 nm). Fig. 3A shows the temperature dependence of the nonlocal spin signal for several typical spacing of 2 μm, 8μm, 14 μm and 20 μm, respectively. Similar to the feature observed on the 10 μm device discussed earlier, the nonlocal spin signals of all the devices exhibit a large enhancement around ~ 20 K, and saturate below the temperature of ~ 5 K. Fig. 3B shows the spacing dependence of the nonlocal spin signals at 2 K and 10 K under the magnetic field of 9 T. Very interestingly, the nonlocal spin signal exhibits a very slow decrease as the spacing increases from 2 μm up to 20 μm, which is quite different from previous results on the ferrimagnetic insulator (YIG), where the nonlocal spin signal exponentially decays as a function of the spacing (*26, 32, 35*). Whereas, in our case for the spin superfluid ground state, the nonlocal spin signal is shown to be inversely proportional to the spacing. This observation agrees well with the theoretical calculation (red dashed lines in Fig. 3B) for spin superfluidity in the presence of the magnetic damping (*22*):

$$R_{NL} = R_{NL}^{d=0} \frac{L_\alpha}{d + L_\alpha} \quad (2)$$

where $R_{NL}^{d=0}$ is related to the spin-dependent chemical potential at the spin injector and the spin mixing conductance ($g^{\uparrow\downarrow}$) at the interfaces between Pt and Cr$_2$O$_3$, and $L_\alpha = \hbar g^{\uparrow\downarrow}/2\pi\alpha S$, where $\hbar$ is the reduced Planck constant, $\alpha$ is the magnetic damping, and $S$ is the saturated spin density in the spin superfluid state. $L_\alpha$ is obtained to be 16.3 ± 1.1 μm at 2 K and 13.3 ± 2.5 μm at 10 K based on the best fitting (red lines in Fig. 3B). As the temperature increases, $L_\alpha$ decreases, and a transition from spin superfluid transport to thermally-generated incoherent magnon transport is observed (*24*). For example, at 80 K, the nonlocal spin signal decays rapidly as a function of spacing, which is very similar to incoherent magnon diffusion in YIG reported previously (*24*).

Since the spin component in the *y-z* plane ($S_y + iS_z$) acquires a free U(1) processional motion of the antiferromagnetic moment within the *y-z* plane, the canted antiferromagnet rotation around the [0001]-direction is more favorable. Hence, we investigate the spin superfluid ground state in the (11$\bar{2}$0)-oriented Cr$_2$O$_3$ films, of which the spins on Cr atoms are aligned along the [0001] orientation and in the film plane. Fig. 4A shows the schematic spin configurations of the (11$\bar{2}$0)-oriented Cr$_2$O$_3$ films epitaxially grown on (11$\bar{2}$0)-Al$_2$O$_3$ (*24*). When the in-plane magnetic field is larger than the spin flop field, a free U(1) processional motion of the antiferromagnetic moment within the *y-z* plane leads to a spin supercurrent of the *x*-component of the spin density, giving rise to the appurtenance of the nonlocal spin signal, as shown in Fig. 4B. The saturation of the nonlocal spin signal at low temperature and the spacing dependence of the nonlocal spin signal at 2 and 10 K are consistent with the spin superfluid model (*22, 24*). Furthermore, as the spin analog of the superconducting effect, spin superfluid ground state exhibits a critical temperature (T$_C$), indicated by the rapid increase of the nonlocal spin signal, as shown in Fig. 4C inset. The T$_C$ vs. the in-plane magnetic field is shown in Fig. 4C. The best fit of the boundary of the spin superfluid phase gives a power law with an exponent of 0.65: $T_C \sim (B-3.5)^{0.65}$, as indicated by the red line. The magnetic field dependence of the associated critical temperature follows a 2/3 power law, suggesting that the transition shares the similar feature as a BEC transition of spin-1 bosons expected theoretically (*10, 13*).

In summary, our experiments demonstrate the spin superfluid ground state in canted antiferromagnetic Cr$_2$O$_3$ thin films at low temperatures. Our discovery of the long distance spin transport in superfluid ground state provides an ideal platform for the fundamental physics on the

BEC of spin-1 bosons and is extremely important for the development of future spin supercurrent-based applications (*15, 36*).

**Acknowledgments:**

We acknowledge the fruitful discussion with Fuchun Zhang, Zhiqiang Wang, Yaroslav Tserkovnyak, and Fa Wang. The research is supported by the financial support from National Basic Research Programs of China (973 program Grant Nos. 2015CB921104, 2015CB921102, 2014CB920902, and 2014CB920901) and National Natural Science Foundation of China (NSFC Grant No. 11574006 and 11534001).

X.C.X. and W.H. proposed and supervised the research. W.Y., T.S., and Y.Y. did the growth of the antiferromagnetic $Cr_2O_3$ thin films, fabricated the spin devices, and performed the nonlocal spin transport measurements. Q.Z., R.S., and X.C.X. performed the theoretical analysis. W.H. wrote the manuscript with contributions from all authors.

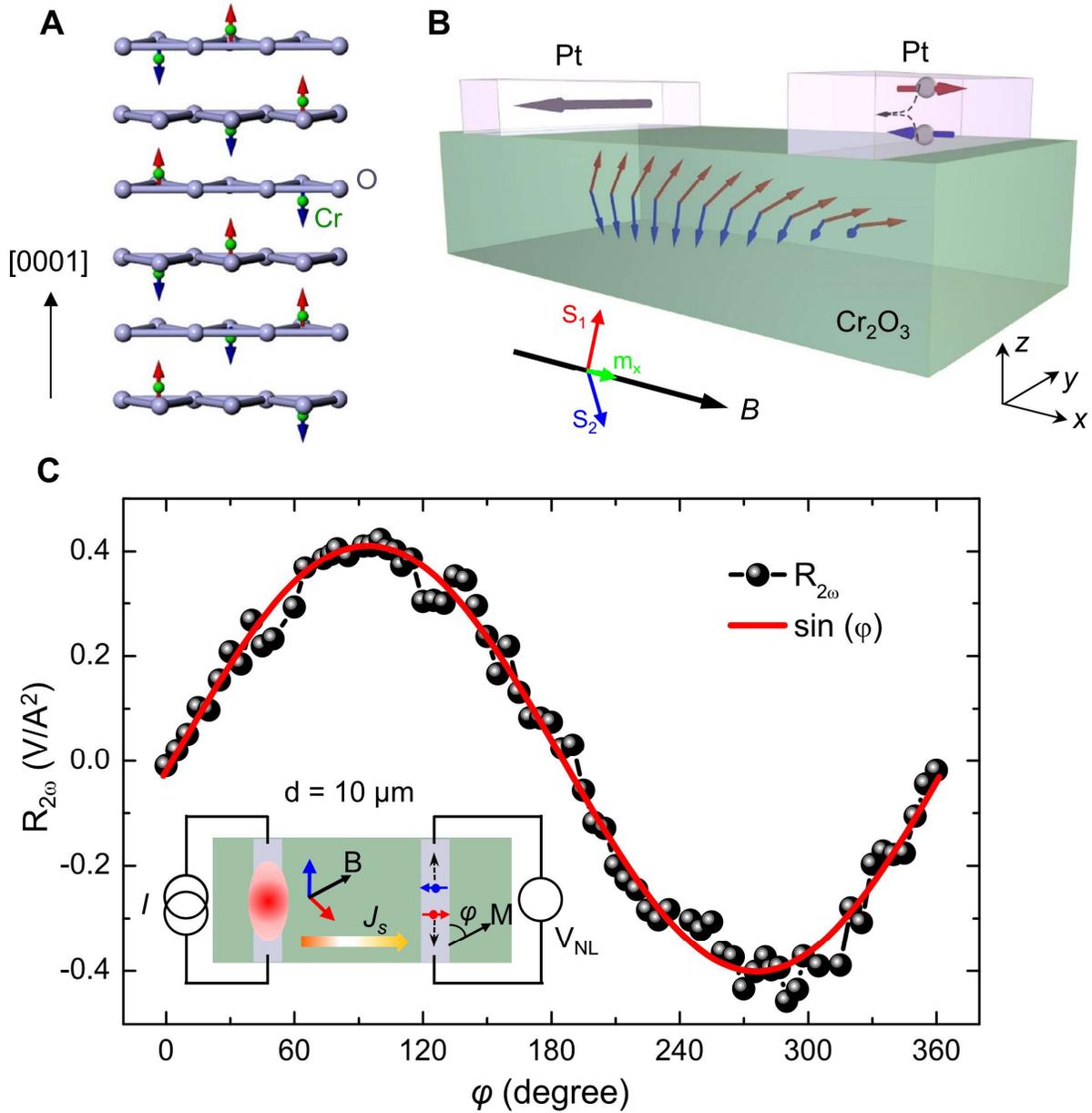

**Fig. 1. The nonlocal spin transport in the spin superfluid ground state of the canted antiferromagnetic $Cr_2O_3$ thin film.** (**A**) The spin structure of single crystalline antiferromagnetic (0001)-oriented $Cr_2O_3$ thin film. Up (red) and down (blue) spins of the $Cr^{3+}$ ions are aligned parallel the crystal's [0001] orientation. (**B**) Schematic of the nonlocal spin transport geometry for the spin transport measurement in the spin superfluid state. The canted magnetization direction is controlled by the external magnetic field ($B$) along the $x$ direction. In such canted antiferromagnetic configuration, the spin component ($S_y + iS_z$) that is perpendicular to the magnetic field direction becomes coherent in the spin superfluid state. (**C**) The second

harmonic resistance in the nonlocal geometry measured on the nonlocal device as a function of the in-plane magnetic field angle at 2 K and 9 T. The spins are injected at the left Pt strip via spin Seebeck effect. The collective spin transport in the spin superfluid ground state is probed at the right Pt strip via inverse spin Hall effect. The nonlocal device is fabricated on the ~ 19 nm $Cr_2O_3$ thin film, and the spacing between two Pt strips ($d$) is 10 μm. The red curve is a sin ($\varphi$) fit for the experimental data (solid balls).

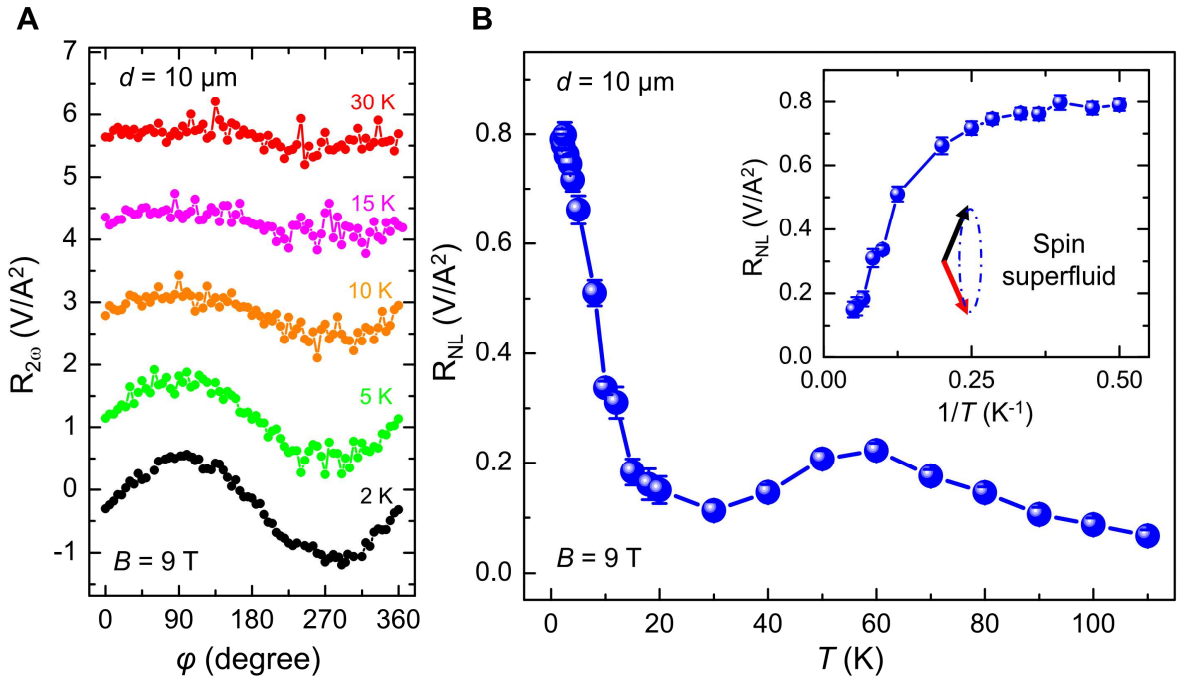

**Fig. 2. Temperature dependence of the nonlocal spin transport in canted antiferromagnetic (0001)-oriented $Cr_2O_3$ thin film.** **(A)** The second harmonic resistance in the nonlocal geometry measured as a function of the in-plane rotation angle under the magnetic field of 9 T at 2, 5, 10, 15, and 30 K, respectively. **(B)** The nonlocal spin signal as a function of the temperature ($T$). Inset: The nonlocal spin signal as a function of $1/T$. At low temperatures in the spin superfluid ground state, the transverse spin component ($S_y + iS_z$) that is perpendicular to the magnetic field direction becomes coherent.

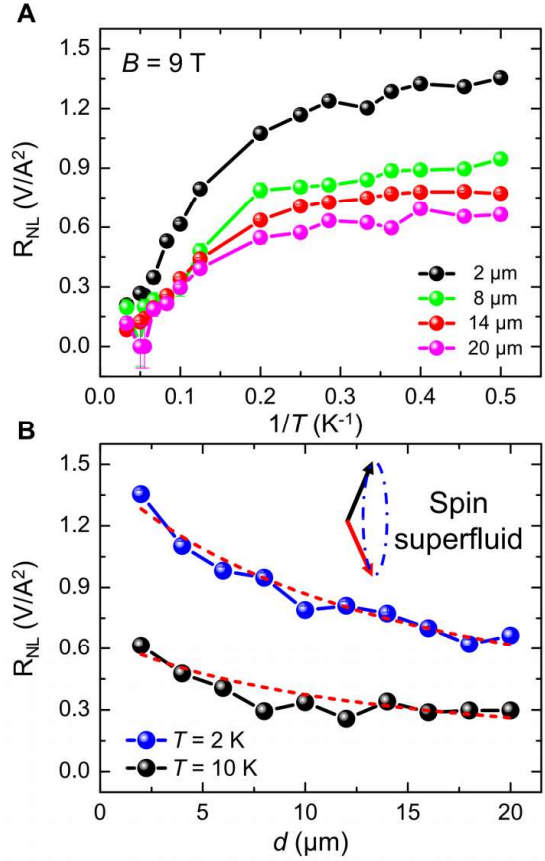

**Fig. 3. Spacing dependence of the nonlocal spin transport in spin superfluid ground state.**
**(A)** The nonlocal spin signal as a function of $1/T$ for the spacing between the two Pt strips (d) of 2 μm, 8 μm, 14 μm, and 20 μm, respectively. These results are obtained under the in-plane magnetic field of 9 T. **(B)** The nonlocal spin signal at 2 K and 10 K in the spin superfluid ground state as a function of the spacing between the two Pt strips. The red dashed lines are the fitting curves based on spin superfluid model using the equation (2).

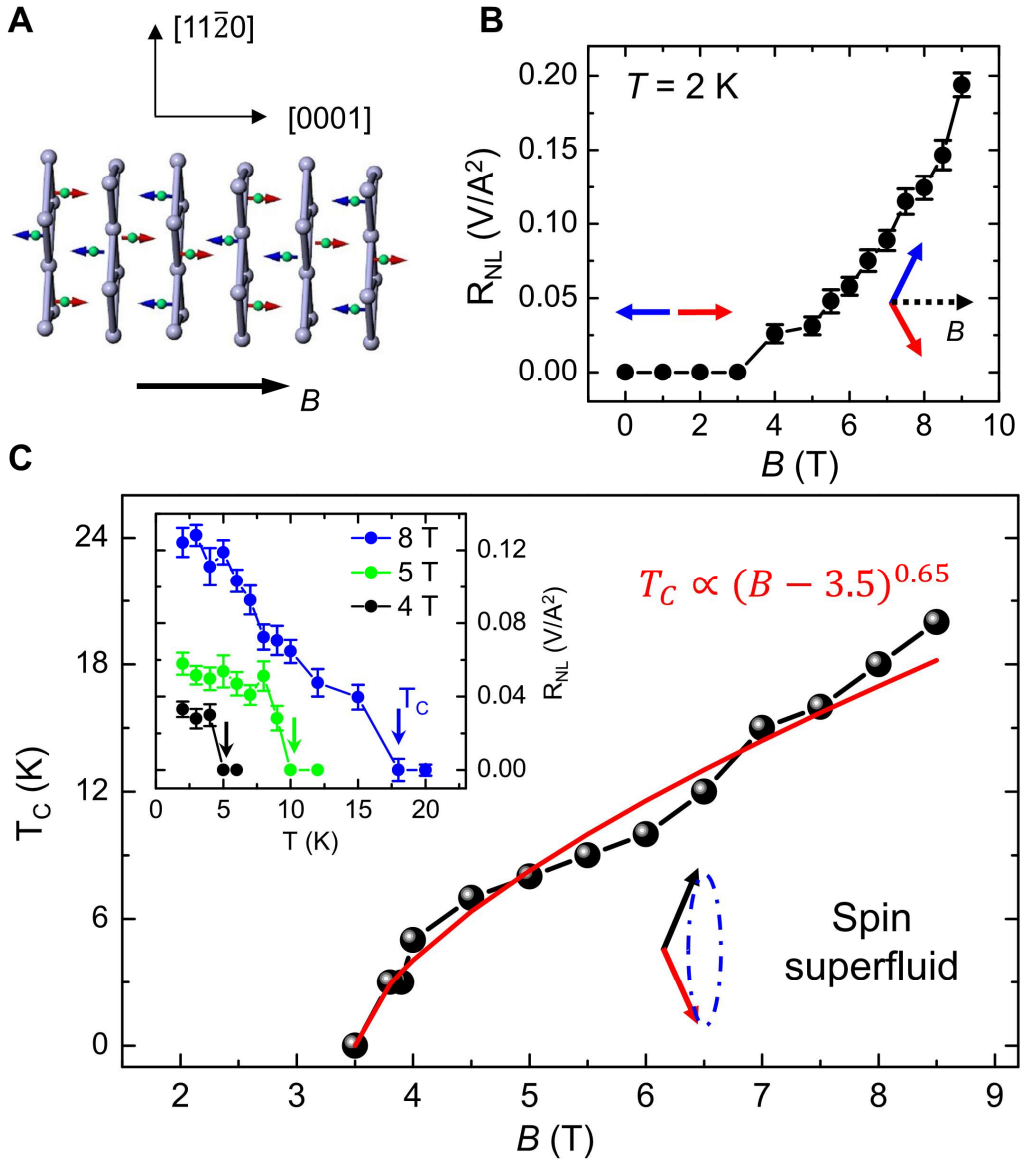

**Fig. 4. Spin superfluid ground state in the canted antiferromagnet (11$\bar{2}$0)-oriented Cr$_2$O$_3$.** **(A)** The spin structure of single crystalline antiferromagnetic (11$\bar{2}$0)-oriented Cr$_2$O$_3$ thin film. Up (red) and down (blue) spins of the Cr$^{3+}$ ions are aligned in the film plane. **(B)** The nonlocal spin signal at 2 K as a function of the magnetic field. The nonlocal spin signal is observed when the magnetic field is higher than the spin flop field. **(C)** The critical temperature as a function of the magnetic field. Red line presents the best fitting curve, which follows a relationship of $T_C \sim (B-3.5)^{0.65}$.

**Supplementary Materials:**

Materials and Methods

Supplementary Text

Figures S1-S15

Full Reference List

# Materials and Methods:

<u>(0001)-oriented $Cr_2O_3$ films growth and characterization.</u> Antiferromagnetic insulating (0001)-oriented $Cr_2O_3$ thin films were epitaxially grown on (0001)-oriented $Al_2O_3$ substrates via pulsed laser deposition (PLD) with a base pressure of $2 \times 10^{-8}$ mbar. Prior to the growth of $Cr_2O_3$ thin films, the $Al_2O_3$ substrates were subsequently cleaned in a mixture ($H_2O_2$:$NH_4OH$:$H_2O$ = 1:1:50) for 1 min, Isopropyl alcohol for 1 min, and DI water for 1min, and then annealed at 1000 °C for 2 hours in an oxygen gas filled furnace. Then the $Al_2O_3$ substrates were loaded into the PLD chamber and the substrate temperature was increased to 500 °C with a rate of 20 °C/min in the oxygen partial pressure of 0.03 mbar. The $Cr_2O_3$ films were deposited from a $Cr_2O_3$ target with a laser power of $(8.0 \pm 0.2)$ mJ and a frequency of 8 Hz. The reflection high energy electron diffraction (RHEED) was used to monitor the growth of antiferromagnetic $Cr_2O_3$ film (Figs. S1A-1B). The $Cr_2O_3$ growth rate was ~ 0.15 nm/min inferred from the oscillations of the RHEED intensity during of the growth. Atomically flat surfaces were achieved, as evidenced by atomic force microscopy (Figs. S1C and S1D). RMS roughness of (0001)-oriented $Al_2O_3$ substrate and ~ 19 nm $Cr_2O_3$ film were 0.12 nm and 0.17 nm, respectively. The thickness of the $Cr_2O_3$ thin film was estimated to be ~ 19 nm from both the growth rate and the X-ray diffraction (XRD) after the film growth (Fig. S1E).

<u>($11\bar{2}0$)-oriented $Cr_2O_3$ films growth and characterization.</u> ($11\bar{2}0$)-oriented $Cr_2O_3$ films were grown on using the ($11\bar{2}0$)-oriented $Al_2O_3$ substrates following the same recipe as (0001)-oriented $Cr_2O_3$ thin films, except that ($11\bar{2}0$)-oriented $Al_2O_3$ substrates were annealed at 1350 °C for 2 hours in an oxygen gas filled furnace prior to the growth. Figs. S2A-2B show the typical RHEED patterns. Atomically flat surfaces were also achieved on ($11\bar{2}0$)-oriented films (Figs.

S2C and S2D), and the RMS roughness of the $Al_2O_3$ substrate and ~ 18 nm $Cr_2O_3$ film were 0.08 nm and 0.14 nm, respectively. Fig. S2E shows the XRD result of this $Cr_2O_3$ film.

Nonlocal device fabrication. The nonlocal devices were fabricated using standard e-beam lithography followed by Pt metal deposition with the growth rate of 0.6 Å /s using DC sputtering system in Ar plasma with a base pressure lower than $8 \times 10^{-7}$ mbar. Then the devices were formed after a lift-off process in Acetone. As shown in Fig. S3, one long Pt strip (width: 300 nm; length: 100 μm) was used for spin injection via the spin Seebeck effect, and other one was used for spin detection via the inverse spin Hall effect. The thickness of the Pt strips is ~ 8 nm, measured by atomic force microscopy. The optical image of the nonlocal device discussed in the main text ($d$ = 10 μm) is shown in the Fig. S3.

Nonlocal spin transport measurements. The nonlocal spin transport measurements were carried out via standard low frequency lock-in technique, as illustrated in Fig. S3. The AC current ($I_{AC}$ = 500 μA unless noted otherwise) with a frequency of 7 Hz was applied using Keithley current source (K6221) and the second harmonic voltage was measured using Signal Recovery lock-in amplifier (SR830). The second harmonic resistance ($R_{2\varpi}$) is calculated from the second harmonic voltage ($V_{2\varpi}$) based on the equation: $R_{2\varpi} = \sqrt{2} V_{2\varpi} / I_{AC}^2$. During the measurement, the nonlocal devices were rotated in-plane under static magnetic fields in the Physical Properties Measurement System (PPMS; Quantum Design). A low noise voltage preamplifier (SR560) was used to enhance the signal-to-noise ratio.

## Supplementary Text

S1. First and Second harmonic results on (0001)-oriented $Cr_2O_3$ thin film

During nonlocal spin transport measurements, two lock-in amplifiers are used to probe the first and second harmonic voltages. The first harmonic resistance ($R_{1\varpi}$) is calculated from the first harmonic voltage ($V_{1\varpi}$) based on the equation: $R_{1\varpi} = V_{1\varpi} / I_{AC}$. And the second harmonic resistance ($R_{2\varpi}$) is calculated from the second harmonic voltage ($V_{2\varpi}$) based on the equation: $R_{2\varpi} = \sqrt{2} V_{2\varpi} / I_{AC}^2$. Fig. S4 shows the typical results measured on two devices ($d$ = 10 μm and 2 μm) at $B$ = 9 T and $T$ = 2 K. For both devices, the second harmonic resistances could be clearly observed, which follow a sin ($\varphi$) relationship as a function of the magnetic field angle (*26*). Whileas, no obvious first harmonic signal is observed within a noise level of 0.05 m$\Omega$.

This observation could be attributed to that the spin injection efficiency by thermal means is much higher than that via spin Hall effect, which is not surprising since it has already been observed in previous studies of spin injection from Pt to ferromagnetic insulators (*32, 33*). Nevertheless, to fully understand the absence of the first harmonic nonlocal spin signal, further theoretical and experimental studies are needed.

S2. Comparison of low temperature and high temperature results on (0001)-oriented Cr$_2$O$_3$ thin film

The low temperature spin transport via spin superfluid ground state has exhibited two major differences compared to the high temperature spin transport via incoherent magnons, as discussed below.

The first major difference is contrasting spacing dependence of the nonlocal spin signal. As stated in the main text, the nonlocal signal decays quite slowly and quantitatively agrees well with the spin superfluid model: $R_{NL} = R_{NL}^{d=0} \frac{L_\alpha}{d + L_\alpha}$ (*22*). At high temperatures, the nonlocal spin signal decays rapidly as the spacing increases. As shown in Fig. S5A, as the spacing increases, the nonlocal spin signal follows a relationship of $R_{NL} \sim \frac{\exp(d/\lambda)}{1-\exp(2d/\lambda)} \sim \frac{1}{d}$ (when $d$ is smaller than the magnon diffusion length ($\lambda$)). This feature is the similar to the diffusion of thermally-generated incoherent magnons in YIG (*26, 32*). The transition from spin superfluid transport to thermally-generated incoherent magnon transport can also be evidenced from the rapid decrease of $L_\alpha$ as the temperature increases above ~ 20 K. When the spacing is larger than $L_\alpha$, the damping is very critical to forbidden the long range spin superfluid transport (*22*).

Besides, the current dependence of the nonlocal spin signals are different for low and high temperatures, as shown in Fig. S6. For low temperature results at 2 and 10 K, a critical current is observed, as indicated in Fig. S6B. However, for the nonlocal spin transport at 80 K, $V_{2\varpi}$ is linearly proportional to $I_{AC}^2$ without a critical current density (Fig. S6D). The critical current observed at low temperatures for the spin superfluid ground state could be attributed to the critical spin torque that is needed to overcome the anisotropy energy to make the canted spins rotate along the magnetic field direction. As discussed in previous theoretical studies that the anisotropy in antiferromagnet defines a critical spin current density at the injector to overcome the antiferromagnet pining (*22*).

## S3. Mechanism for the modest enhancement of the nonlocal spin signal at ~ 60 K

The modest enhancement of the nonlocal spin signal on the ~ 19 nm (0001)-oriented thin film is observed at ~ 60 K. To our best knowledge, it could be due to two mechanisms inferred from previous studies. The first one is due to the large spin fluctuation near the Neel temperature of the $Cr_2O_3$ thin film (*28-30*). And the second one is related to magnon-phonon drag effect (*37, 38*). To probe the mechanism, we experimentally measure the Neel temperature of the ~19 nm $Cr_2O_3$ thin film via exchange bias and its thermal conductivity via the 3ω method (*39, 40*). Fig. S7 shows the exchange bias between the 2 nm Py and $Cr_2O_3$ thin films as a function of the temperature. A blocking temperature of ~ 260 K is observed, which ruled out the mechanism of large spin fluctuation to account for our observation. For the thermal conductivity measurements on the (0001)-oriented $Cr_2O_3$ bulk single crystal and ~19 nm thin films, a typical device is shown in Fig. S8A, where the electrodes are made of Pt. The thermal conductivity results of the bulk $Cr_2O_3$ (Fig. S8B) are similar to the previous study using the heat capacity method (*25*). Fig. S8C shows the thermal conductivity of the ~ 19 nm thin film as a function of the temperature. A modest peak is observed at ~ 70 K. It seems that that the magnon-phonon coupling could be the possible cause for the modest enhancement of the nonlocal spin signal at ~ 60 K. However, to fully understand this high temperature modest enhancement, further theoretical and experimental investigations would be essential.

## S4. Supporting results on extra samples: 6 nm and 45 nm (0001)-oriented $Cr_2O_3$ films

Extra samples are measured to further confirm the spin transport via spin superfluid ground state probed on the ~19 nm (0001)-oriented $Cr_2O_3$ film. Fig. S9 and S10 show the nonlocal spin transport results measured on the ~ 6 nm and ~ 45 nm (0001)-oriented $Cr_2O_3$ films, respectively.

Similar to ~19 nm (0001)-oriented $Cr_2O_3$ film, the rapid enhancement and saturation of the nonlocal spin signal at low temperatures are observed (Fig. S9A and S10A), consistent with theoretical predications (*15*). At low temperatures, the slow decaying of the spin signal as the spacing increases (Fig. S9B and S10B), which is consistent with the spin transport via spin superfluid model (*22*). Besides, as the temperature increases, the transition from spin superfluid transport to thermally-generated incoherent magnon diffusion is observed (Fig. S9C, S9D, S10C, and S10D), similar to the results on ~19 nm (0001)-oriented $Cr_2O_3$ film.

Fig. S11 shows the nonlocal spin signal measured on all the (0001)-oriented $Cr_2O_3$ films with various thicknesses. The nonlocal spin signal is larger for thicker films, which is due to higher spin superfluid densities in thicker films.

## S5. First and Second harmonic results on ($11\bar{2}0$)-oriented $Cr_2O_3$ thin film

Fig. S12 shows the typical results measured on two devices fabricated on the ~ 18 nm (11-20)-oriented $Cr_2O_3$ thin film ($d$ = 10 and 2 μm) at $B$ = 9 T and $T$ = 2 K. For both devices, the second harmonic resistances could be clearly observed, which follow a sin ($\varphi$) relationship as a function of the magnetic field angle. Whileas, no obvious first harmonic signal is observed within a noise level of 0.02 mΩ. This observation is similar to the results on (0001)-oriented $Cr_2O_3$ thin films.

## S6. Temperature and spacing dependence of the nonlocal spin transport on ($11\bar{2}0$)-oriented $Cr_2O_3$ thin film

Fig. S13 shows the nonlocal spin transport results measured on the ~ 18 nm ($11\bar{2}0$)-oriented $Cr_2O_3$ thin film at $B$ = 9 T. The rapid enhancement and saturation of the nonlocal spin signal at low temperatures are observed (Fig. S13A), which is consistent with the spin superfluid transport expected theoretically (*15*). The spacing dependence of the nonlocal spin signal at $T$ = 2 and 10 K (Fig. S13B) is also consistent with the spin superfluid model (red dashed lines) (*22*), but very different from the exponential decaying expected for thermally-generated incoherent magnons (green dashed lines).

Different from the low temperature spin transport via spin superfluid, the spin transport via incoherent magnons is observed at high temperatures (Fig. S13C). The transition from spin superfluid transport to thermally-generated incoherent magnon transport can also be evidenced from the rapid decrease of $L_\alpha$ as the temperature increases (Fig. S13D). Besides, the current dependence of the nonlocal spin signals are also different at low and high temperatures, as shown in Fig. S14. For low temperature results at 2 and 10 K, a critical current is observed (Fig. S14A and S14B). This is different from the nonlocal spin transport at 80 K (Fig. S14C and S14D) with no signatures of critical current.

S7. Magnetic field dependence of the nonlocal spin transport on (0001)-oriented $Cr_2O_3$ thin film

The role of canted angle for the spin transport in the spin superfluid state is studied for the ~ 19 nm (0001)-oriented $Cr_2O_3$ film. The canted angles can be controlled by the magnetic field strength. Fig. S15A shows the nonlocal vs. the magnetic field angle under 1, 3, 5, 7, and 9 T on the nonlocal device ($d$ = 10 µm) at 2 K. These magnetic fields give rise to very tiny canted angles since they are significantly smaller compared to the saturation magnetic field. The nonlocal spin signal exhibits a linear relationship vs. the in-plane magnetic field (Fig. S15B). This is because the total spin polarization at the detector is proportional to the product of the spin superfluid density and the field-induced canted magnetization along $x$ direction ($m_x$). For all of the magnetic fields from 1 T to 9 T, the nonlocal spin signal exhibits similar features (Fig. S15C), including the large enhancement from ~ 20 K, and the saturation when the temperature is below ~ 5 K. To study temperature dependence of the spin superfluid density, we normalize the nonlocal spin signal by the value at 2 K (Fig. S15D). The normalized nonlocal spin signals for all the magnetic fields almost fall on the same curve, indicating a negligible effect of the magnetic field strength on the saturated spin density for the spin superfluid ground state.

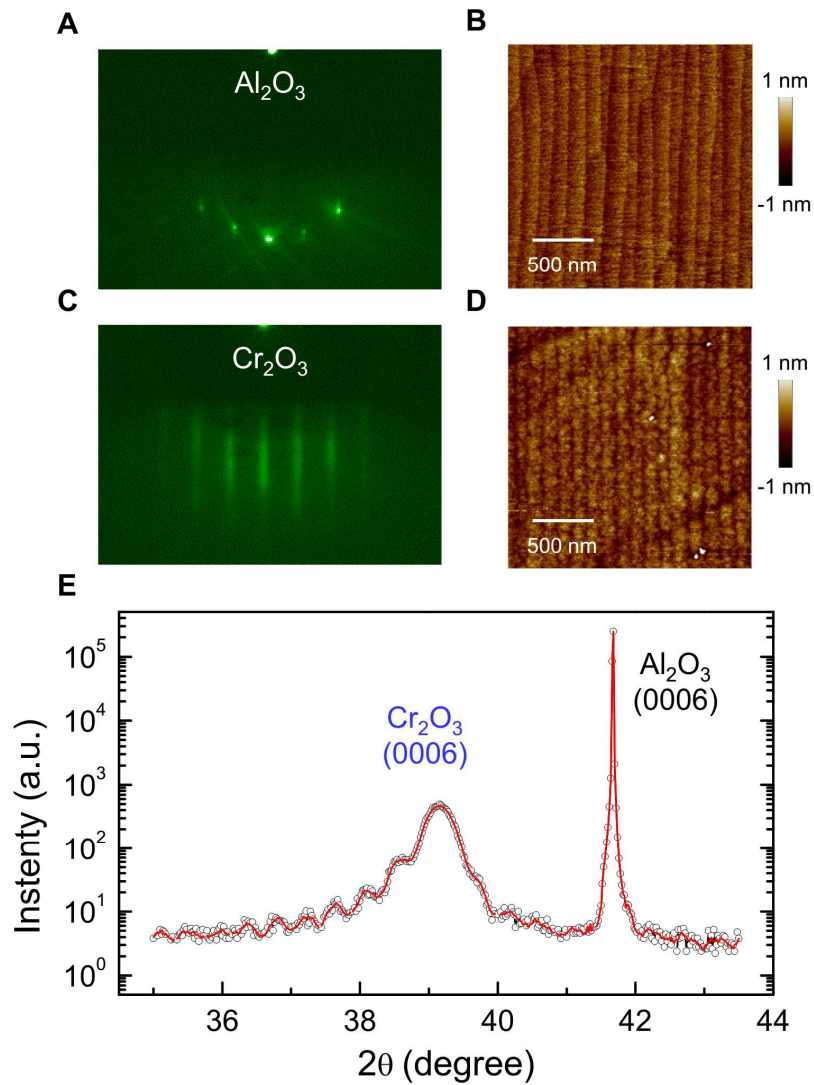

**Fig. S1. RHEED and XRD of the (0001)-oriented $Cr_2O_3$ thin film (~ 19 nm) on the (0001)-oriented $Al_2O_3$ substrate.** **(A-D)** RHEED patterns and the AFM results of the (0001)-oriented $Al_2O_3$ substrate and the ~ 19 nm epitaxial (0001)-oriented $Cr_2O_3$ thin film. The width of the atomic steps is ~ 100 nm. **(E)** High resolution XRD pattern of the ~ 19 nm epitaxial (0001)-oriented $Cr_2O_3$ thin film on the $Al_2O_3$ substrate. The thickness of $Cr_2O_3$ thin film could be determined by the fitting curve (red line).

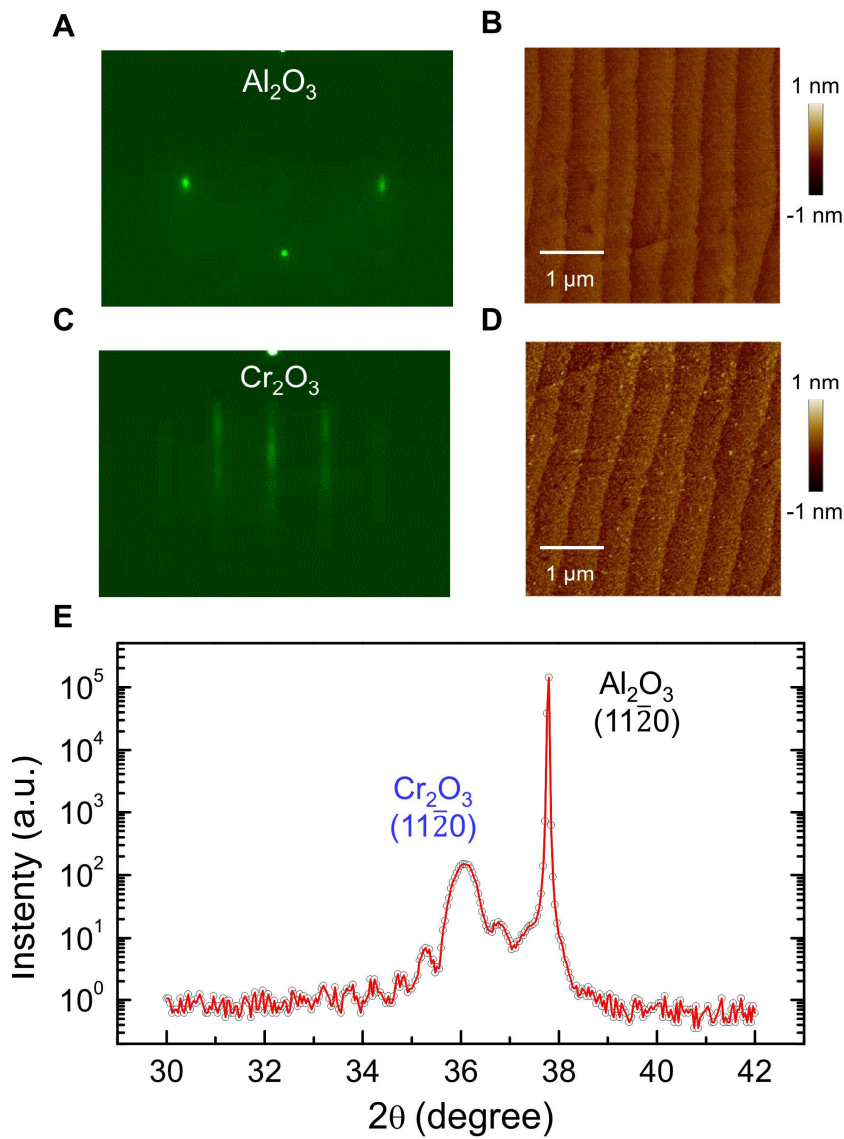

**Fig. S2. RHEED and XRD of the (11$\bar{2}$0) -oriented Cr$_2$O$_3$ thin film (~ 18 nm) on the (11$\bar{2}$0)-oriented Al$_2$O$_3$ substrate.** **(A-D)** RHEED patterns and the AFM results of the (11$\bar{2}$0)-oriented Al$_2$O$_3$ substrate and the ~ 18 nm epitaxial (11$\bar{2}$0)-oriented Cr$_2$O$_3$ thin film. The width of the atomic steps is ~ 1.0 μm. **(E)** High resolution XRD pattern of the ~ 18 nm epitaxial (11$\bar{2}$0)-oriented Cr$_2$O$_3$ thin film on the Al$_2$O$_3$ substrate. The thickness of Cr$_2$O$_3$ thin film could be determined by the fitting curve (red line).

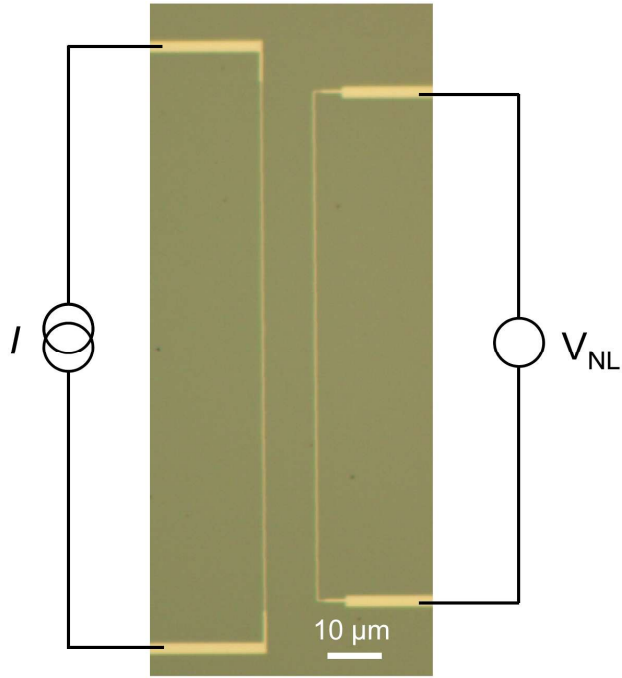

**Fig. S3. The typical optical image of the nonlocal device and the nonlocal measurement.** The thickness of the (0001)-oriented $Cr_2O_3$ is ~ 19 nm, and the spacing between two Pt strips (the spin injector and detector) is 10 μm. The nonlocal spin signal is measured in the nonlocal geometry with standard low frequency lock-in technique.

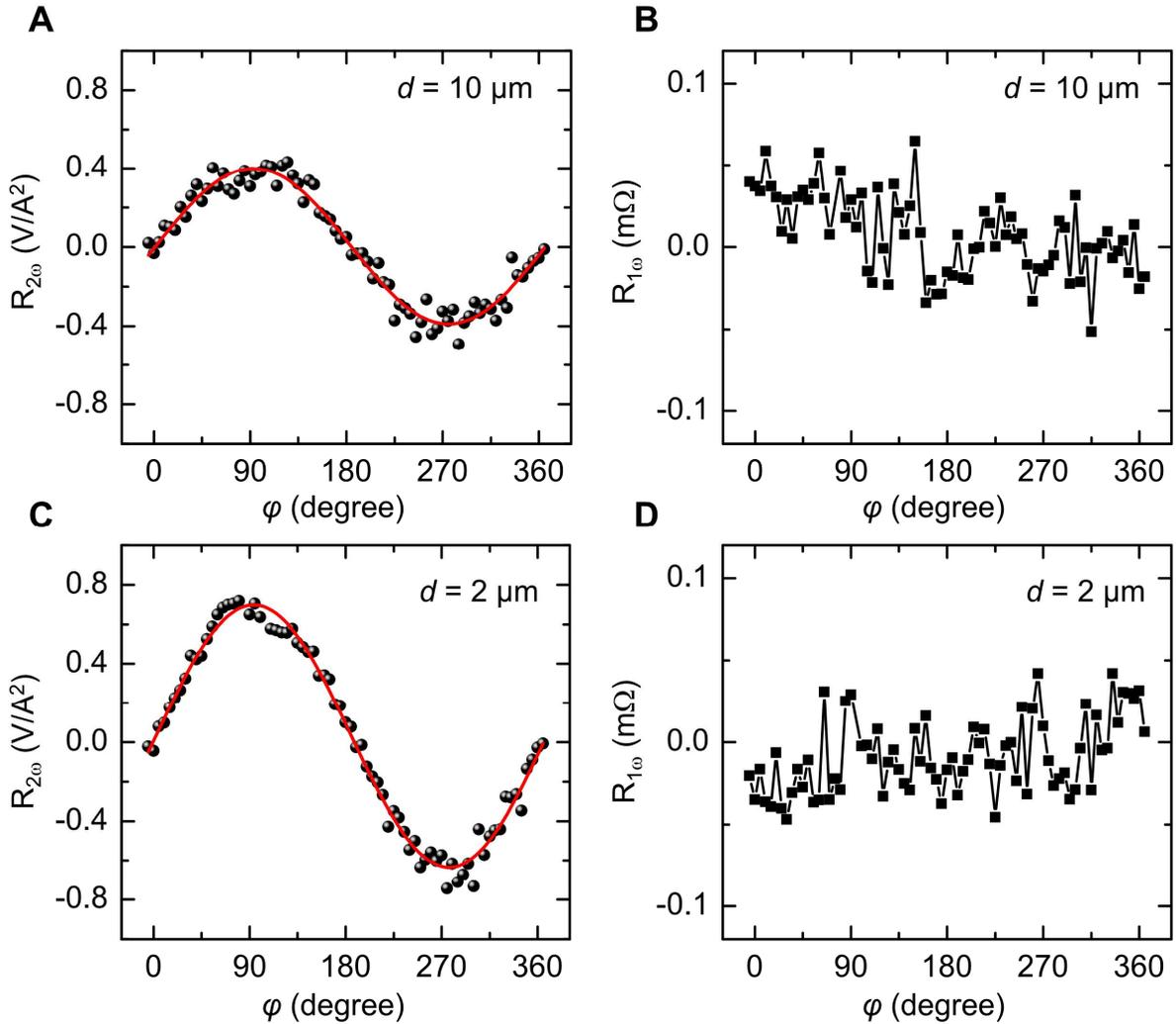

**Fig. S4. The first and second harmonic nonlocal resistance on the ~ 19 nm (0001)-oriented $Cr_2O_3$ film. (A-B)** The first and second harmonic nonlocal resistance for the device of $d$ = 10 μm at $B$ = 9 T and $T$ = 2 K. **(C-D)** The first and second harmonic nonlocal resistance for the device of $d$ = 2 μm at $B$ = 9 T and $T$ = 2 K. The second harmonic nonlocal resistance is proportional to sin ($\varphi$), indicated by red lines in **A** and **C**. No clear first harmonic nonlocal spin signal is observed for both devices.

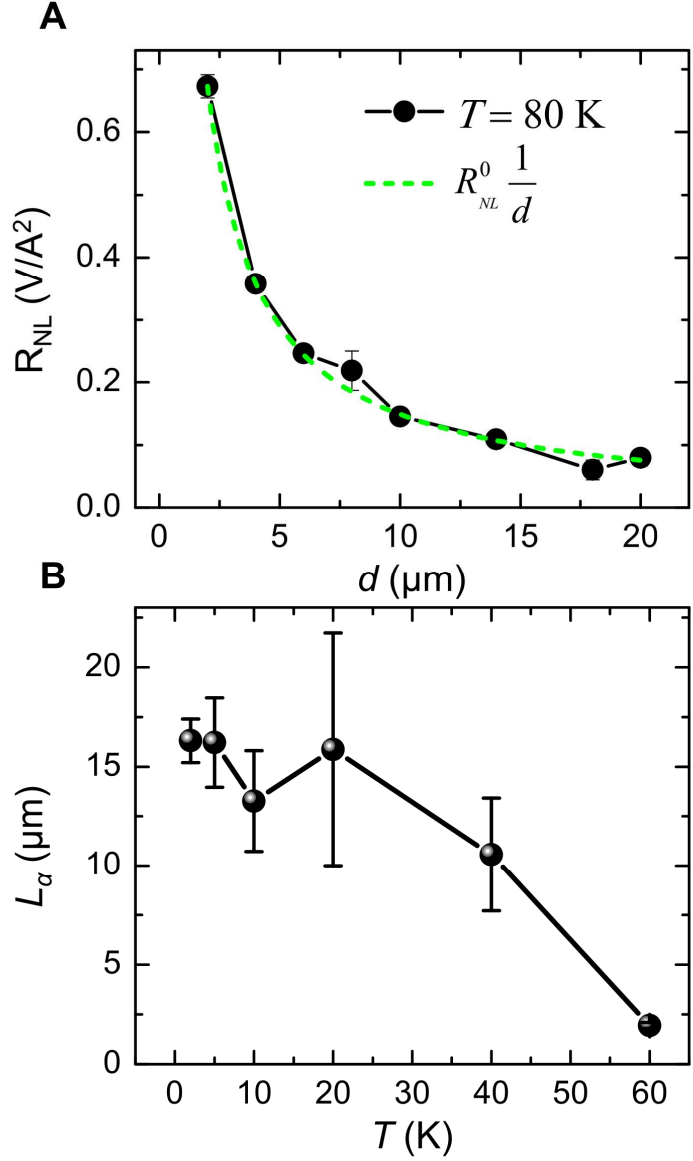

**Fig. S5. The transition from spin transport via spin superfluid to that via thermally-generated incoherent magnons.** **(A)** The spacing dependence of the nonlocal spin signal at 80 K obtained on the ~ 19 nm (0001)-oriented $Cr_2O_3$ film. The green dashed line is the fitting curve based on the incoherent magnon diffusion model ($R_{NL} \sim \frac{1}{d}$). **(B)** The temperature dependence of $L_\alpha$ obtained based on the spin superfluid transport model ($R_{NL} \sim \frac{1}{d + L_\alpha}$).

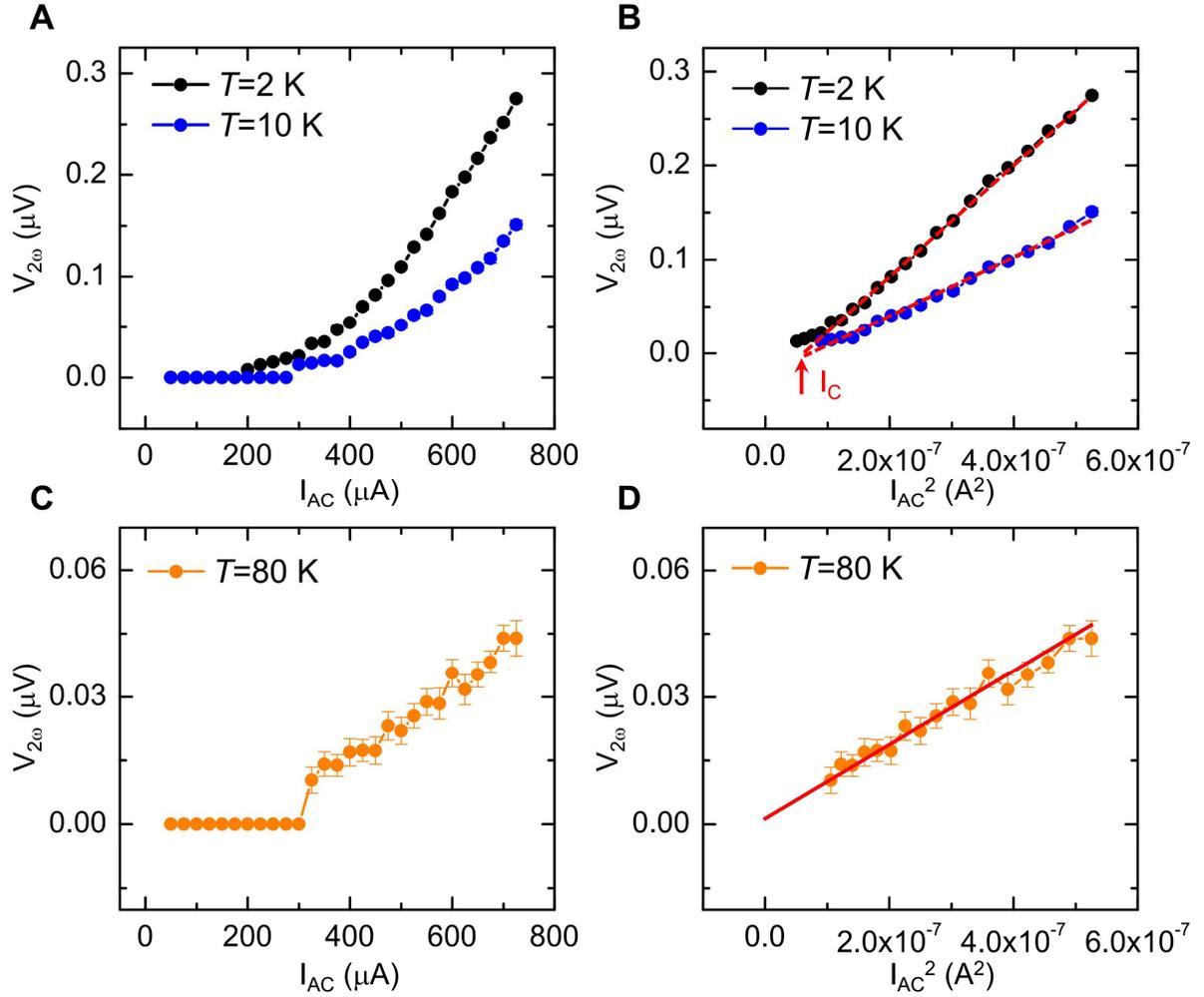

**Fig. S6. Current dependence of the nonlocal spin transport on the ~ 19 nm (0001)-oriented Cr$_2$O$_3$ film. (A-B)** The second harmonic spin voltage vs. $I$ and $I^2$ at $T$ = 2 and 10 K and $B$ = 9 T on the device with $d$ = 10 μm. A critical current ($I_C$) is observed, which is needed to overcome uniaxial anisotropy to induce the spin superfluid transport. **(C-D)** The second harmonic spin voltage vs. $I$ and $I^2$ at $T$ = 80 K and $B$ = 9 T on the device with $d$ = 10 μm. The second harmonic voltage is proportional to $I^2$ without a critical current.

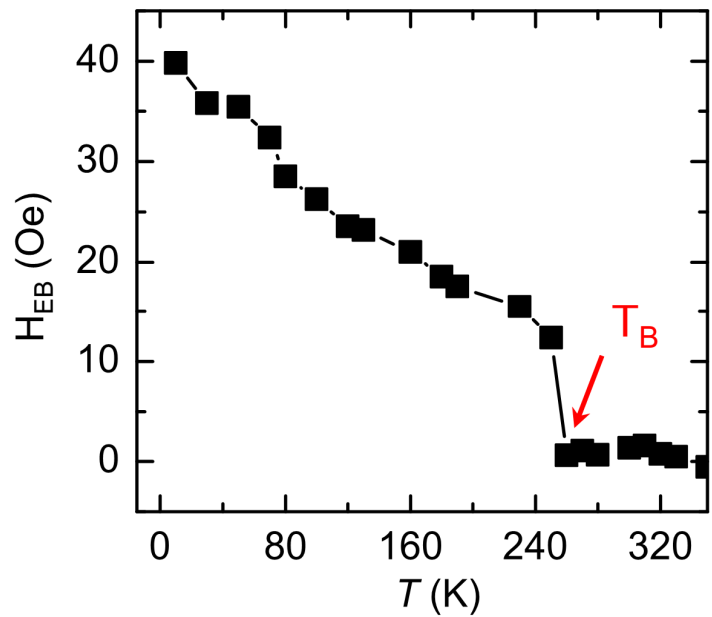

**Fig. S7. The exchange bias between 2 nm Py and the ~ 19 nm (0001)-oriented film.** The blocking temperature is ~ 260 K.

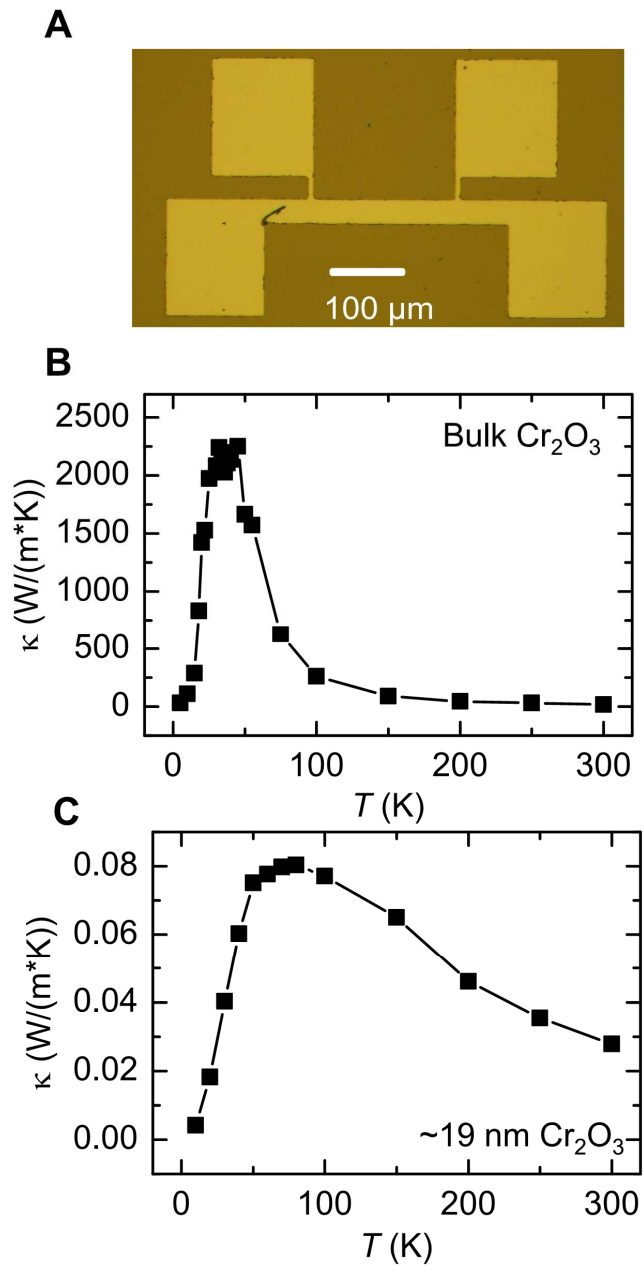

**Fig. S8. Thermal conductivity (κ) measured by 3ω method for (0001)-oriented $Cr_2O_3$ film.** **(A)** The optical image of the device fabricated on the film, where the Pt width is 40 μm. **(B-C)** The thermal conductivity of a single crystal of (0001)-oriented bulk $Cr_2O_3$ crystal and the ~ 19 nm $Cr_2O_3$ film, respectively.

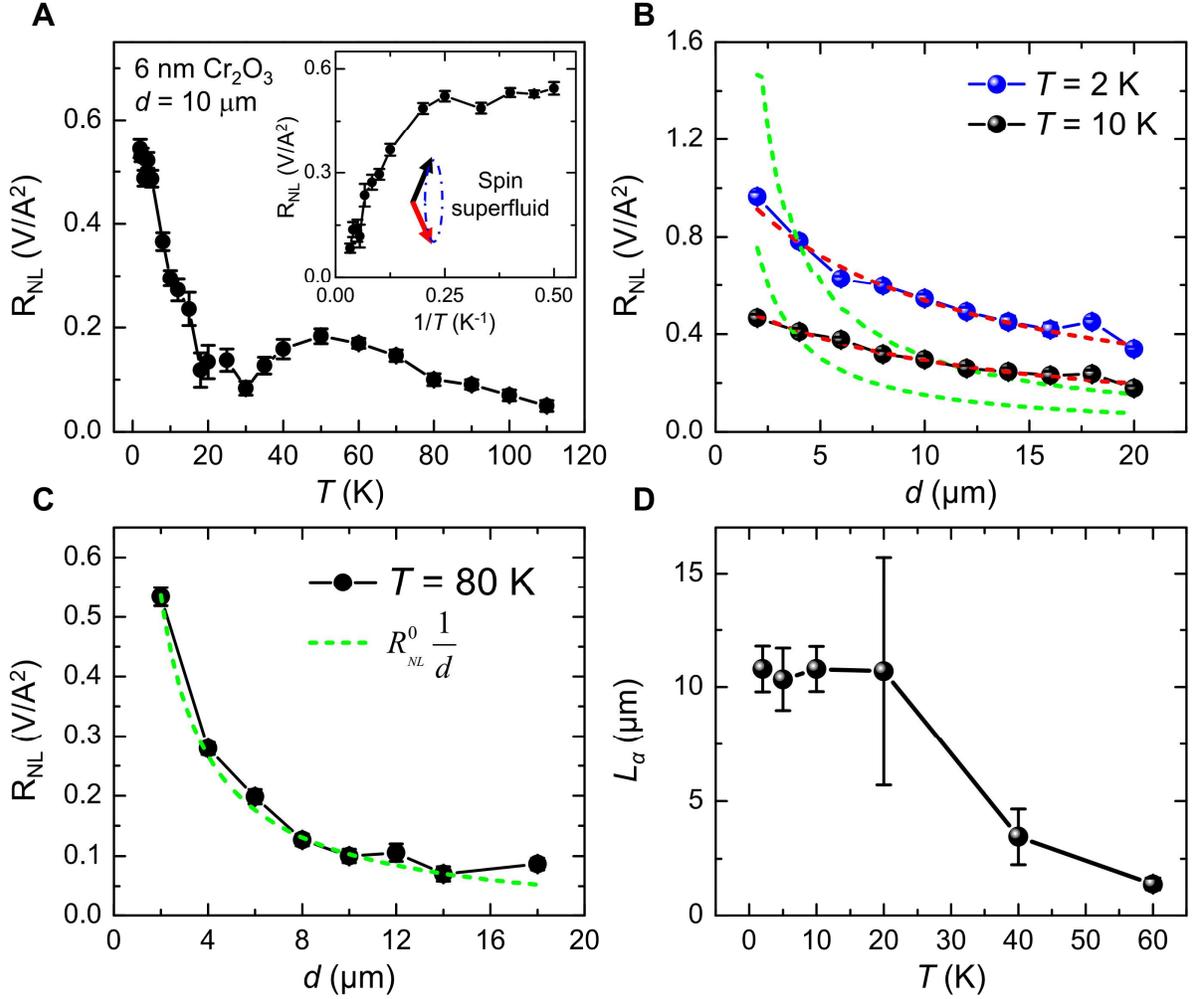

**Fig. S9. The nonlocal spin transport on the ~ 6 nm (0001)-oriented Cr$_2$O$_3$ film. (A)** The temperature dependence of the nonlocal spin signal for the device with $d$ = 10 μm. Inset: The nonlocal spin signal as a function of $1/T$. **(B-C)** The spacing dependence of the nonlocal spin signal at $T$ = 2 K, 10, and 80 K. The red dashed lines are the fitting curves based on spin superfluid model ( $R_{NL} \sim \frac{1}{d+L_\alpha}$ ), and the green dashed lines are fitting curves based on the incoherent magnon diffusion model ( $R_{NL} \sim \frac{1}{d}$ ). **(D)** The temperature dependence of $L_\alpha$ obtained based on the spin superfluid transport model.

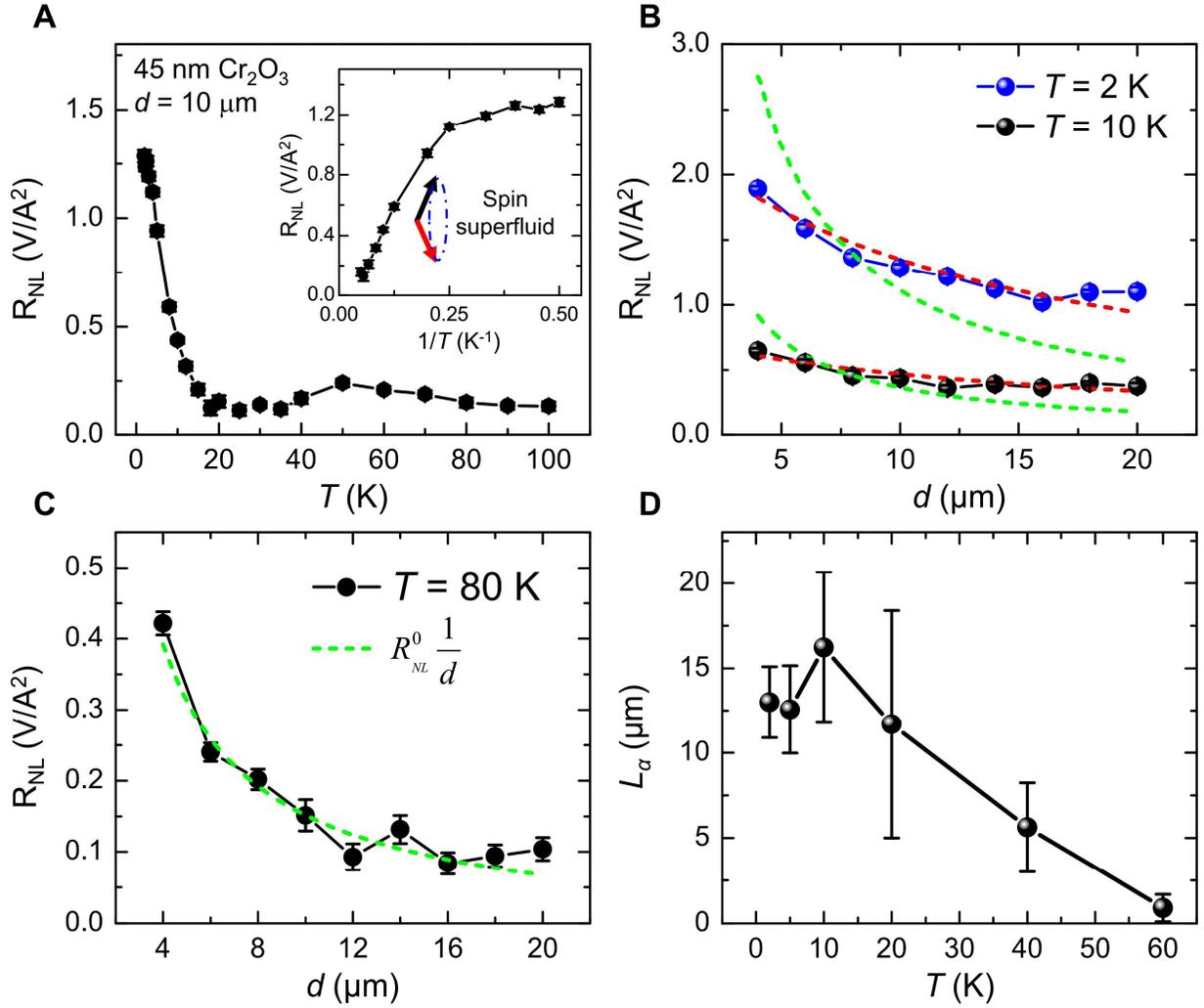

**Fig. S10. The nonlocal spin transport on the ~ 45 nm (0001)-oriented Cr$_2$O$_3$ film.** (**A**) The temperature dependence of the nonlocal spin signal for the device with $d$ = 10 μm. Inset: The nonlocal spin signal as a function of $1/T$. (**B-C**) The spacing dependence of the nonlocal spin signal at $T$ = 2 K, 10, and 80 K. The red dashed lines are the fitting curves based on spin superfluid model ($R_{NL} \sim \frac{1}{d+L_\alpha}$), and the green dashed lines are fitting curves based on the incoherent magnon diffusion model ($R_{NL} \sim \frac{1}{d}$). (**D**) The temperature dependence of $L_\alpha$ obtained based on the spin superfluid transport model.

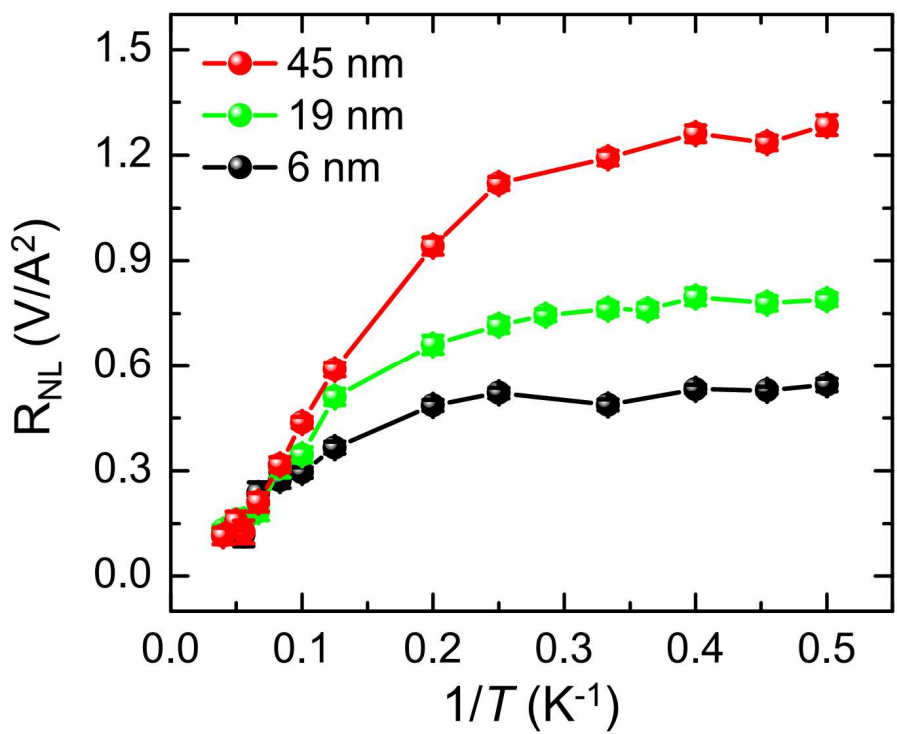

**Fig. S11. Temperature dependence of the nonlocal spin transport for (0001)-oriented $Cr_2O_3$ films with various thicknesses.** $R_{nonlocal}$ vs. $1/T$ for the $d = 10$ μm devices fabricated on 6 nm, 19 nm, and 45 nm thick (0001)-oriented $Cr_2O_3$ films at $B = 9$ T.

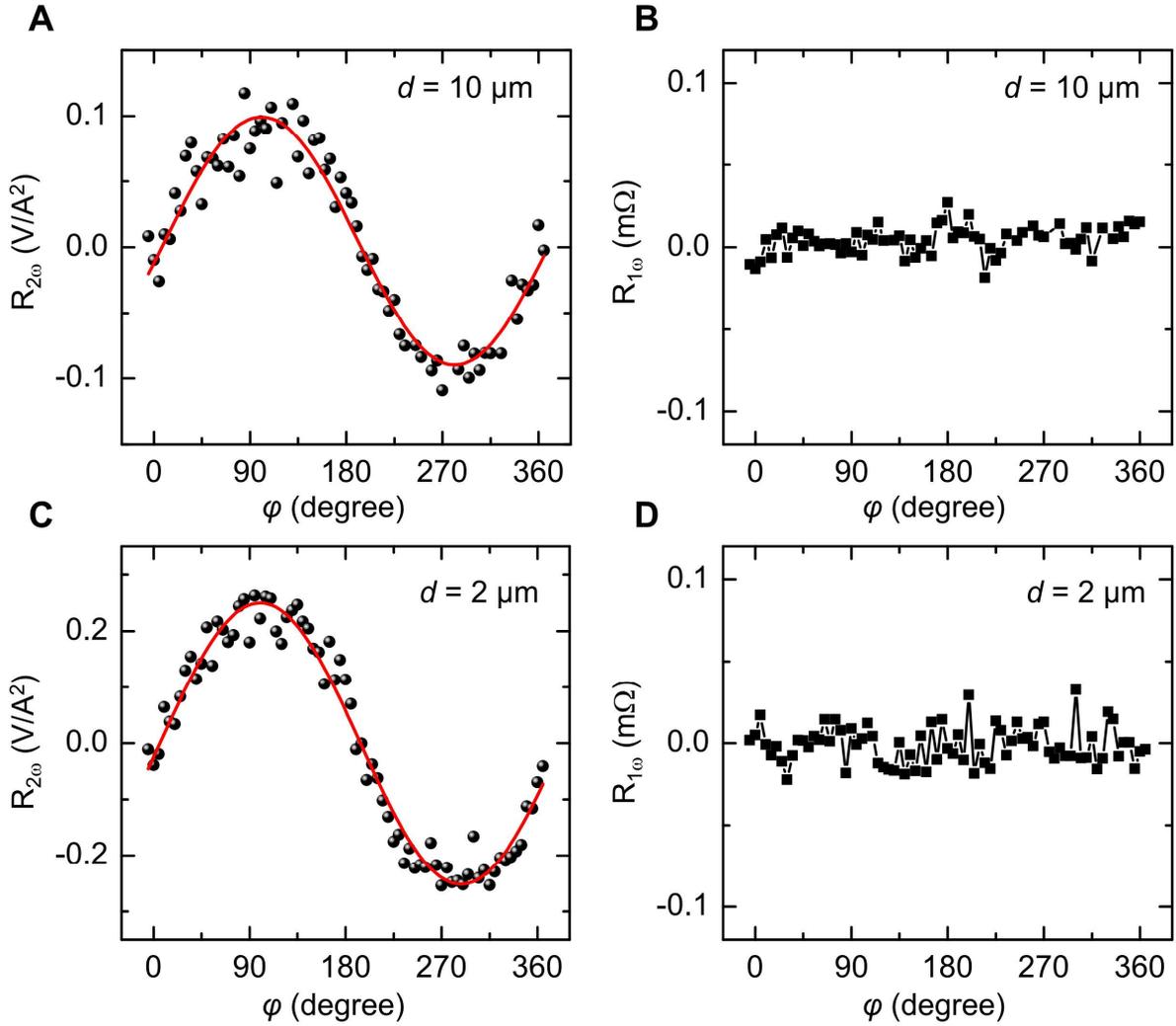

**Fig. S12. The first and second harmonic nonlocal resistance on the ~ 18 nm ($11\bar{2}0$)-oriented $Cr_2O_3$ film.** **(A-B)** The first and second harmonic nonlocal resistance for the device of $d$ = 10 μm at $B$ = 9 T and $T$ = 2 K. **(C-D)** The first and second harmonic nonlocal resistance for the device of $d$ = 2 μm at $B$ = 9 T and $T$ = 2 K. The second harmonic nonlocal resistance is proportional to sin ($\varphi$), indicated by red lines in **A** and **C**. No clear first harmonic nonlocal spin signal is observed for both devices.

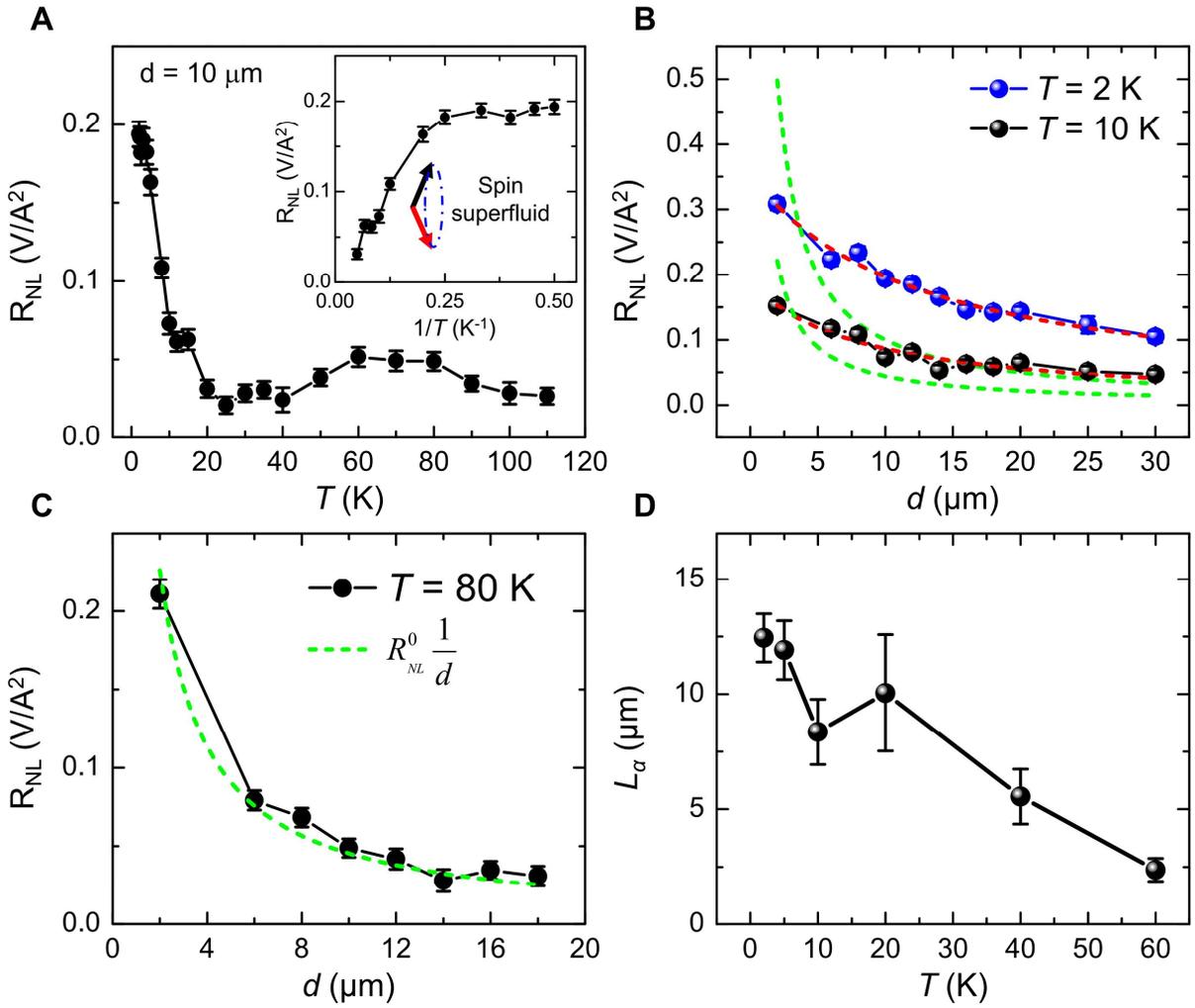

**Fig. S13. The nonlocal spin transport on the ~ 18 nm (11$\bar{2}$0)-oriented Cr$_2$O$_3$ film. (A)** The temperature dependence of the nonlocal spin signal for the device with $d$ = 10 μm at $B$ = 9 T. Inset: The nonlocal spin signal as a function of $1/T$. **(B-C)** The spacing dependence of the nonlocal spin signal at $T$ = 2 K, 10, and 80 K. The red dashed lines are the fitting curves based on spin superfluid model ($R_{NL} \sim \dfrac{1}{d+L_\alpha}$), and the green dashed lines are fitting curves based on the incoherent magnon diffusion model ($R_{NL} \sim \dfrac{1}{d}$). **(D)** The temperature dependence of $L_\alpha$ obtained based on the spin superfluid transport model.

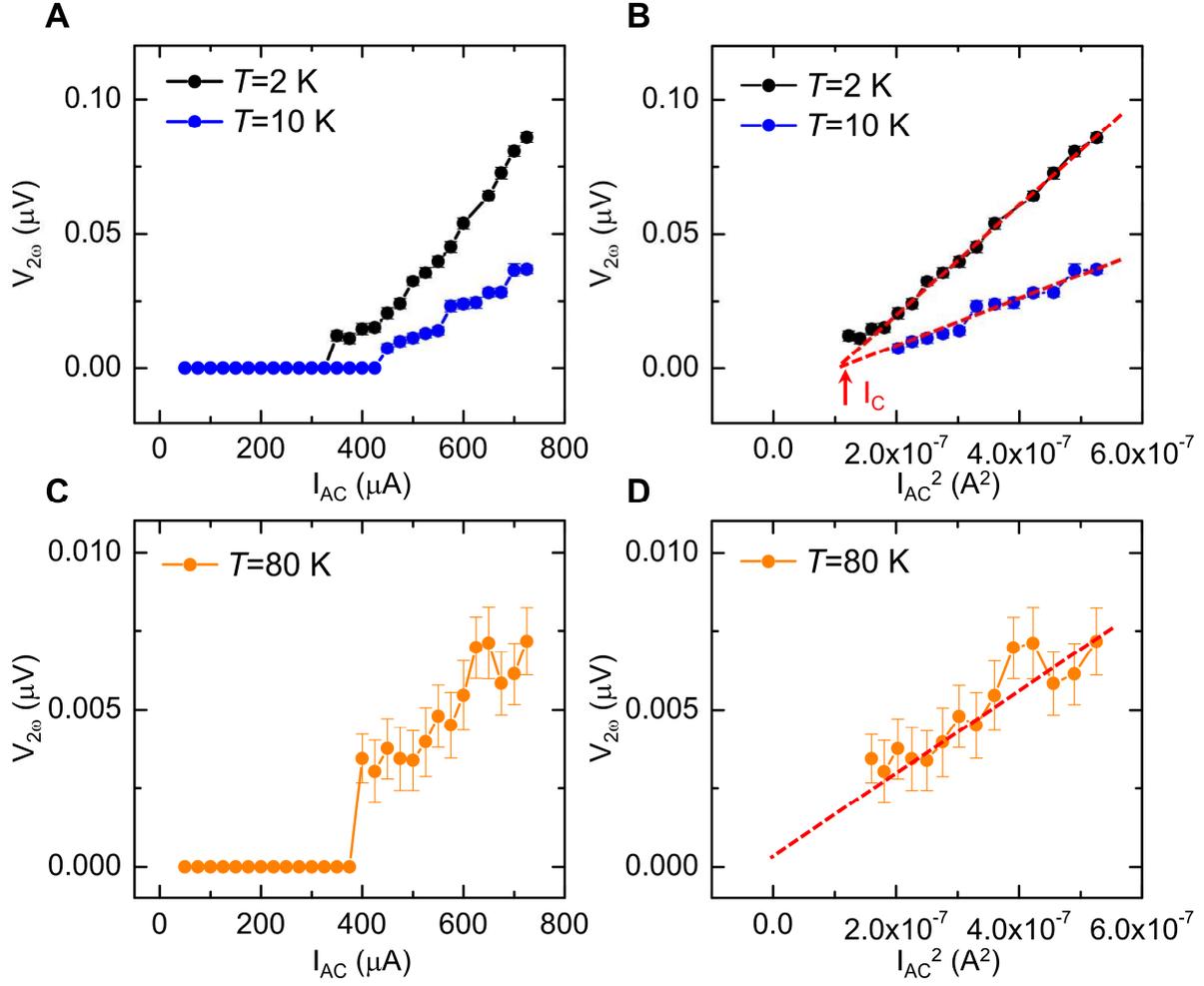

**Fig. S14. Current dependence of the nonlocal spin transport on the ~ 18 nm ($11\bar{2}0$)-oriented $Cr_2O_3$ film. (A-B)** The second harmonic spin voltage vs. $I$ and $I^2$ at $T$ = 2 and 10 K and $B$ = 9 T on the device with $d$ = 10 μm. A critical current ($I_C$) is observed, which is needed to overcome uniaxial anisotropy to induce the spin superfluid transport. **(C-D)** The second harmonic spin voltage vs. $I$ and $I^2$ at $T$ = 80 K and $B$ = 9 T on the device with $d$ = 10 μm. The second harmonic voltage is proportional to $I^2$ without a critical current.

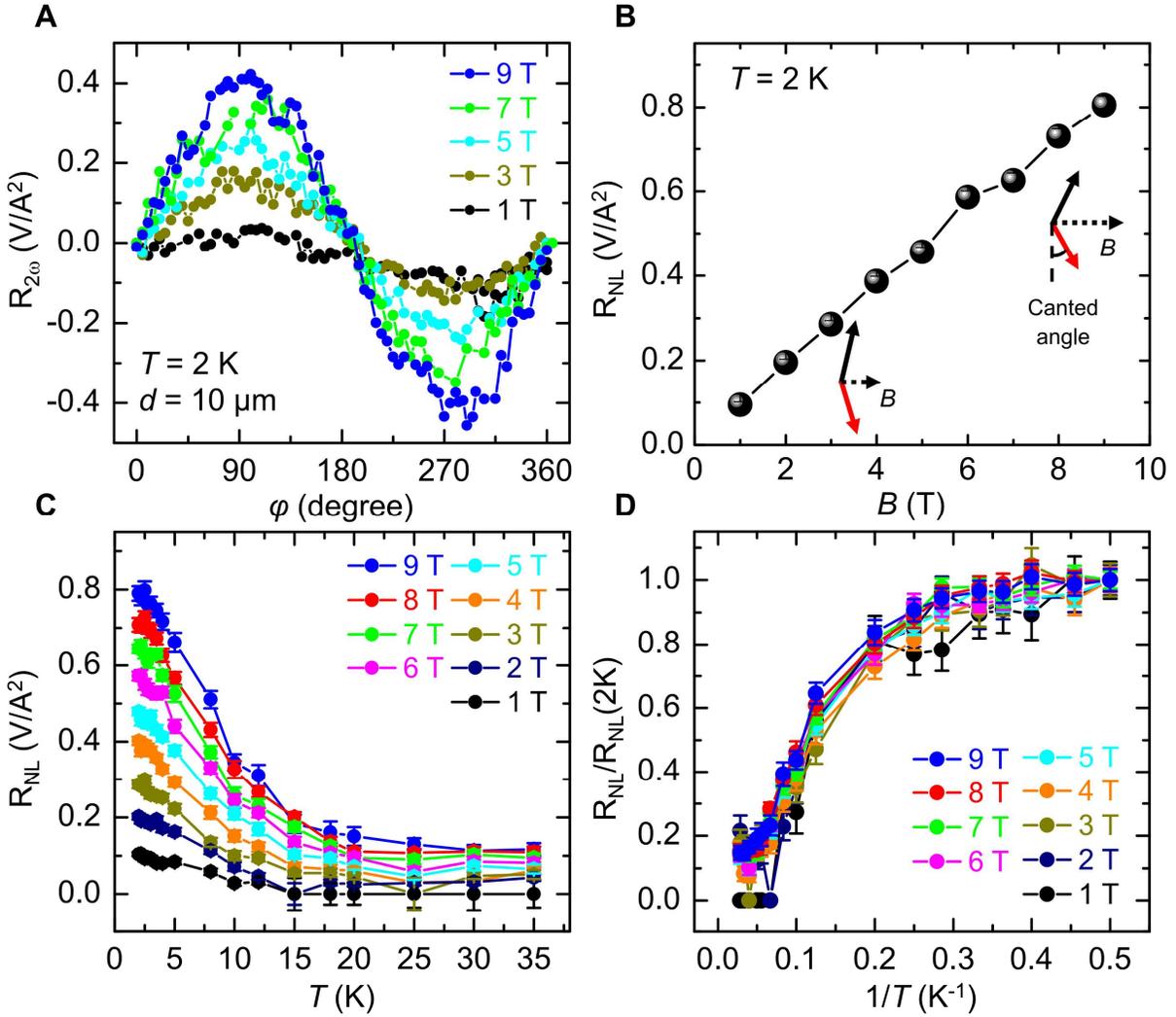

**Fig. S15. Magnetic field dependence of nonlocal spin transport on the ~ 19 nm (0001)-oriented $Cr_2O_3$ film.** **(A)** The nonlocal second harmonic resistance as a function of the magnetic field angle at 2 K under the magnetic fields of 1 T, 3 T, 5 T, 7 T, and 9 T, respectively. **(B)** The nonlocal spin signal at 2 K as a function of the magnetic field. A larger magnetic field corresponds to a larger canted angle of the two antiferromagnetic spins. **(C)** The nonlocal spin signal as a function of the temperature under the magnetic fields from 1 T to 9 T. **(D)** The normalized nonlocal spin signal ($R_{NL}/R_{NL}(2K)$) as a function of $1/T$ under the magnetic fields from 1 T to 9 T.